\documentclass{aa}  

\usepackage{graphicx}
\usepackage{txfonts}
\usepackage{amsmath}
\usepackage{amssymb}
\usepackage{adjustbox}
\usepackage[colorlinks=true,linkcolor=blue,citecolor=blue,urlcolor=blue]{hyperref}%

\newcommand{\lya}{Ly{$\rm \alpha$}\xspace}
\newcommand{\jwst}{{\em JWST}\xspace}
\newcommand{\hi}{\ion{H}{i}}
\newcommand{\hii}{\ion{H}{ii}}

\newcommand{\oiii}{\ion{O}{iii}}
\newcommand{\oii}{\ion{O}{ii}}
\newcommand{\neiii}{\ion{Ne}{iii}}
\newcommand{\hei}{\ion{He}{i}}

\newcommand{\civ}{\ion{C}{iv}}
\newcommand{\ciii}{\ion{C}{iii}}
\newcommand{\heii}{\ion{He}{ii}}
\newcommand{\oi}{\ion{O}{i}}
\begin{document}

   \title{Uncovering the physical origin of the prominent Lyman-$\alpha$ emission and absorption in GS9422 at $z = 5.943$}


   \author{Chamilla~Terp,\inst{1,2}
          \and
          Kasper~E.~Heintz,\inst{1,2,3}
          \and
          Darach~Watson,\inst{1,2}
          \and
          Gabriel~Brammer,\inst{1,2}
          \and
          Adam~Carnall,\inst{4}
          \and
          Joris~Witstok,\inst{5,6}
          \and
          Renske~Smit,\inst{7}
          \and 
          Simone~Vejlgaard,\inst{1,2}
          }
    \institute{
    Cosmic Dawn Center (DAWN), Denmark 
    \and
    Niels Bohr Institute, University of Copenhagen, Jagtvej 128, 2200 Copenhagen N, Denmark 
    \and
    Department of Astronomy, University of Geneva, Chemin Pegasi 51, 1290 Versoix, Switzerland 
    \and
    Institute for Astronomy, University of Edinburgh, Royal Observatory, Edinburgh EH9 3HJ, UK  
    \and
    Kavli Institute for Cosmology, University of Cambridge, Madingley Road, Cambridge CB3 0HA, UK 
    \and
    Cavendish Laboratory, University of Cambridge, 19 JJ Thomson Avenue, Cambridge CB3 0HE, UK 
    \and
    Astrophysics Research Institute, Liverpool John Moores University, Liverpool, L35 UG, UK  
    }

   \date{Received \today; accepted \today}

 
  \abstract
   {We present a comprehensive spectro-photometric analysis of the galaxy GS9422 from the JADES GTO survey located at $z=5.943$, anomalously showing a simultaneous strong \lya\ emission feature and damped \lya\ absorption (DLA), based on \jwst NIRSpec and NIRCam observations. The best-fit modelling of the spectral energy distribution (SED) reveals a young, low-mass (${\rm log}(M_\star/M_{\odot}) = 7.8 \pm 0.01$) galaxy, with a mass-weighted mean age of the stellar population of $(10.9^{+0.07}_{-0.12})\,$Myr. The identified strong nebular emission lines suggest a highly ionized ($O_{32} = 59$), low-metallicity ($12+\log({\rm O/H}) = 7.78\pm 0.10$) star-forming galaxy with a star-formation rate SFR = ($8.2 \pm 2.8$) $\rm M_{\odot}\;yr^{-1}$ over a compact surface area $A_e = 1.85$\,kpc$^{2}$, typical for galaxies at this epoch. This corresponds to an intense SFR surface density of $\log (\Sigma_{\rm SFR} / M_\odot\,{\rm yr}^{-1}\,{\rm kpc}^{-2}) = 1.14\pm 0.30$. We carefully model the rest-frame UV NIRSpec Prism spectrum around the \lya\ edge, finding that the \lya\ emission-line redshift is consistent with the longer-wavelength recombination lines and an escape fraction of $f_{\rm esc,Ly\alpha} = 30\%$ but that the broad DLA feature is not able to converge on the same redshift. Instead, our modelling suggests $z_{\rm abs}= 5.40 \pm 0.10$, the exact redshift of a newly identified proto-cluster in nearby projection to the target galaxy. We argue that most of the \hi\ gas producing the strong \lya\ damping wing indeed has to be unassociated with the galaxy itself, and thus may indicate that we are probing the cold, dense circumcluster medium of this massive galaxy overdensity. These results provide an alternative solution to the recent claims of continuum nebular emission or an obscured active galactic nucleus dominating the rest-frame UV parts of the spectrum and provide further indications that strong DLAs might preferentially be associated with galaxy overdensities. 
   }

   \keywords{high-redshift galaxies
               }

    \titlerunning{The physical origin of the prominent Lyman-$\alpha$ emission and absorption in a galaxy at $z = 5.943$}
    \authorrunning{Terp et al.}

   \maketitle
%
\section{Introduction}\label{sec:intro}

The history of the first billion years of cosmic time accounts the transition from primordial, neutral hydrogen and helium atoms into the first stars and galaxies, the synthesis of heavier elements in stellar cores, and the eventual reionization of the Universe. With the launch and powerful near-infrared capabilities of the {\em James Webb Space Telescope} (\jwst), we are now able to study and constrain these processes within this critical era of the early Universe \citep{Robertson22}. Spectroscopic observations with \jwst/NIRSpec \citep{Jakobsen22} have in particular been paramount in spectroscopically identifying the most distant galaxies to date at $z\approx 11-13$ \citep{CurtisLake23,Wang23b,Fujimoto23_uncover,Bunker23_gnz11} and charting the chemical enrichment of galaxies at $z>6$ \citep[e.g.,][]{Schaerer22,ArellanoCordova22,Taylor22,Brinchmann23,Curti23a,Katz23,Rhoads23,Trump23,Heintz23_JWSTALMA,Nakajima23,Langeroodi23,Sanders24}. These early observations revealed that galaxies at $z>7$ appear chemically ``diluted'' based on the negative offset of their gas-phase metallicities from the otherwise fundamental-metallicity relation \citep{Heintz23_FMR}, though the exact redshift for this transition is still debated \citep{Nakajima23,Curti23}. This indicates that \jwst is starting to uncover the formation phase of galaxies, at a point where they are still intimately connected to the intergalactic medium (IGM) and experiencing excessive \hi\ gas overflow. 

Indeed, a significant fraction of galaxies at $z>8$ have been discovered with extremely strong damped Lyman-$\alpha$ (\lya) absorption \citep[DLA;][see also \citealt{Umeda23,DEugenio23}]{Heintz23_DLA}, in excess of the damping wings expected for a largely neutral intergalactic medium at these redshifts \citep{MiraldaEscude98,McQuinn08,Keating23}. These observations imply large \hi\ gas column densities $N_{\rm HI}\gtrsim 10^{22}\,$cm$^{-2}$ and covering fractions, and may be prevalent in $\gtrsim 65\%$ of the galaxy population at $z>8$ \citep{Heintz24} At lower redshifts, near the end of reionization at $z\approx 6$, the fraction of galaxies showing strong integrated DLAs decrease to $\approx 30\%$ and only represents the youngest star-forming systems that are yet to substantially ionize their surrounding gas or process most of the neutral, atomic hydrogen gas into molecules and stars. 

As part of the \jwst/NIRSpec PRImordial gas Mass AssembLy (PRIMAL) survey \citep[\jwst-PRIMAL;][]{Heintz24} targeting galaxies during the reionization epoch at $z>5.5$ with robust spectroscopic redshifts and continuum sensitivity near \lya, we identified one intriguing source showing strong (\lya) emission {\em and} damped \lya\ absorption at $z=5.94$. Intuitively, this seems to counter the general observed trends and strains the physical interpretation of substantial \hi\ gas producing strong \lya\ damping wings but at the same time enabling the escape of \lya\ photons. Here we thus aim to characterize the source in detail and provide some physical scenarios that might explain these extreme observables. We note that this particular case has also been studied by \citet{Cameron_2023}, who present evidence for the rest-frame UV shape of this galaxy spectrum being dominated by nebular two-photon emission. However, the high stellar temperatures required ($100,000\,$K), the observed weak rest-optical emission lines and ionizing photon production efficiency, and relatively high gas-phase metallicity make it ambiguous \citep{Chen23}. Alternatively, this source has been proposed to host an obscured active galactic nucleus (AGN) where the rest-UV light reflect a central young stellar disk \citep{Tacchella24,Li24}. This thus motivates a more detailed analysis of this particular source.  

We have structured the paper as follows. In Sect.~\ref{sec:obs} we describe the photometric and spectroscopic observations and in Sect.~\ref{sec:res} we present our analysis and results. In Sect.~\ref{sec:disc} we place the characteristics of the source into context, classify the likely underlying source of emission, and present an alternative scenario to resolve this rare conundrum of a simultaneous strong (\lya) emission {\em and} DLA feature observed in the integrated galaxy spectrum. Throughout the paper we assume concordance flat $\Lambda$CDM cosmology, with $H_0 = 67.4$\,km\,s$^{-1}$\,Mpc$^{-1}$, $\Omega_{\rm m} = 0.315$, and $\Omega_{\Lambda} = 0.685$ \citep{Planck18}.

\section{JWST Observations}\label{sec:obs}

The source (named GS9422 hereafter) was observed with \jwst/NIRSpec \citep{Jakobsen22} as part of the GTO \jwst Advanced Deep Extragalactic Survey (JADES, prog. ID: 1210, PI: Luetzendorf) with MSA source ID 13176 as detailed in \citet{Eisenstein23,Bunker23}. The observations were separated into 28 hours integration in the low-resolution Prism/CLEAR and 7 hours integration in each of the medium-resolution gratings G140M/F070LP, G235M/F170LP, and G395M/F290LP. The Prism spectrum covers the entire wavelength range of NIRSpec from $0.7-5.3\mu$m at a resolving power $\mathcal{R}\approx 100$. The NIRSpec grating covers smaller bandwidths but at a spectral resolution of $\mathcal{R}\approx 1000$. We adopt the reduced and processed spectra delivered by the DAWN \jwst Archive (DJA)\footnote{\url{https://dawn-cph.github.io/dja/index.html}} reduced through the custom-made {\sc MSAExp} pipeline \citep{Brammer_msaexp}\footnote{version 0.6.17; DOI:  
10.5281/zenodo.7299500}. We refer to \citet{Heintz23_DLA} for further details of the reduction process. Throughout this work, we mainly consider the NIRSpec Prism observations for the continuum modelling of the spectral energy distribution (SED) and the medium-resolution grating spectra for the emission line fluxes and kinematics. 

We further include \jwst/NIRCam \citep{Rieke23} imaging of the source obtained as part of JADES \citep{Eisenstein23} in the seven broad-band filters: F090W, F115W, F150W, F200W, F277W, F356W, and F444W. We adopt the photometry listed in DJA for all filters, derived in a $0\farcs 5$ circular aperture around the centroid of the source at R.A. (J2000) = $03^{\rm h}32^{\rm m}29.2^{\rm s}$ and Decl. (J2000) = $-27^\circ 47' 51.48''$. The images have been processed using {\sc Grizli} \citep{Brammer_grizli}\footnote{DOI:  
10.5281/zenodo.6672538} which astrometrically calibrates each field to the {\em Gaia}-DR3 reference frame and drizzles the images to a common pixel scale of $0\farcs 04$/pixel \citep[see][for further details]{Valentino23}. We further use photometry to rescale the 1D spectra to account for potential slit losses and improve the absolute flux calibration of the spectra. 

\begin{figure*}
    \centering
    \includegraphics[width=6cm]{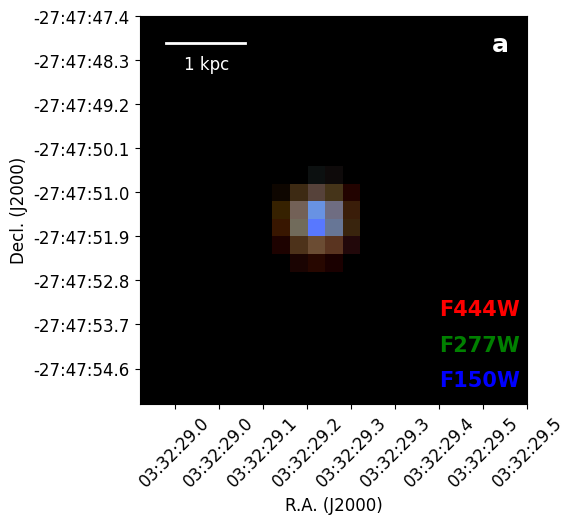}
    \includegraphics[width=11cm]{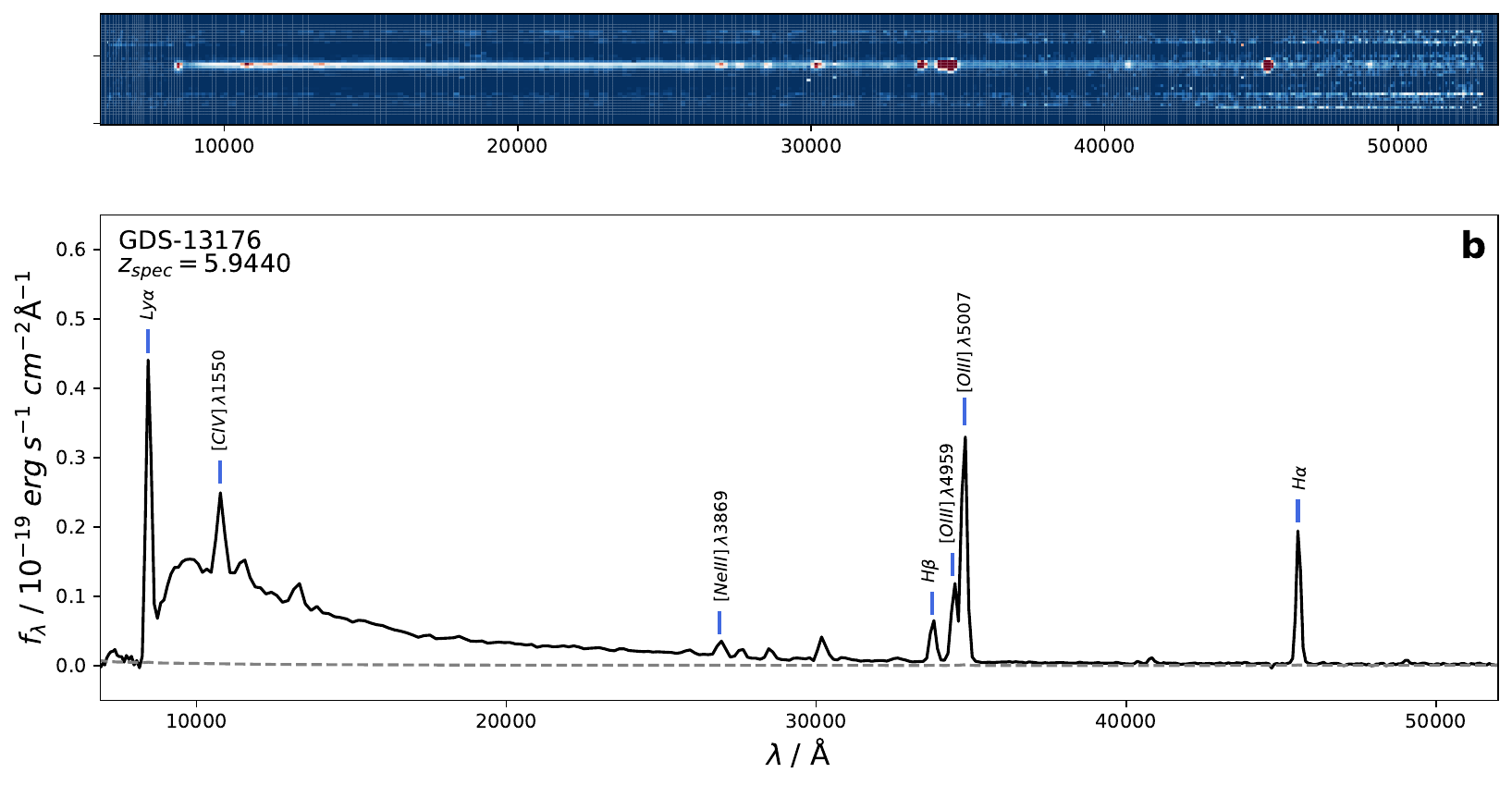}
    \caption{\textbf{(a)} False-RGB-plot zoomed in on GS9422 based on the \jwst/NIRCam filters: F444W (red), F277W (green) and F150W (blue). The physical scale is calculated in the source plane. \textbf{(b)} \jwst/NIRSpec Prism spectrum with the most prominent nebular and auroral emission lines marked. The top panel shows the 2D trace spectrum. }
    \label{fig:introfig}
\end{figure*}

\section{Analysis and results} \label{sec:res}

\subsection{Redshift and line fluxes}

\begin{figure}
    \centering
    \includegraphics[width=8cm]{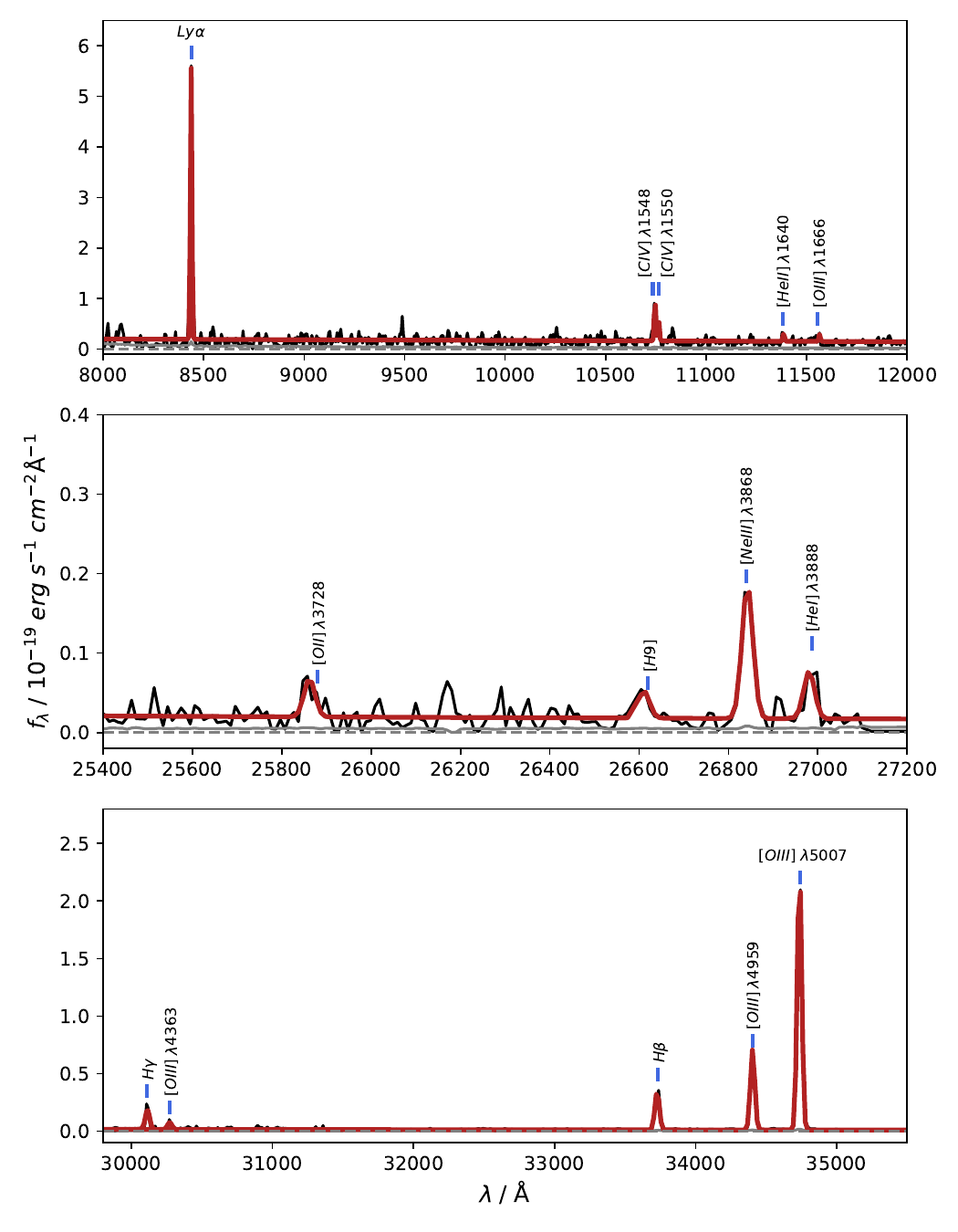}
    \caption{Zoomed-in views of the regions of medium-resolution gratings with the main nebular and auroral emission line transitions marked. The best-fit continuum and Gaussian profiles to each line transition are shown by the solid red line.}
    \label{fig:med_res_fit}
\end{figure}

Based on the extracted 1D \jwst/NIRSpec medium-resolution grating spectra, we detect and derive the line fluxes of the following nebular and auroral emission lines: \lya, [\civ]\,$\lambda \lambda 1548, 1550$, [\heii]\,$\lambda$1640, [\oiii]\,$\lambda$1666, [\oii]\,$\lambda\lambda 3727,3729$, H9, [\neiii]$\lambda$3869, [\hei]$\lambda$3889, [\oiii]$\lambda$4363, 4959 and 5007 as well as the Balmer lines, H$\alpha$ and H$\beta$ (some of these lines are also detected in the Prism spectrum). We further detect and derive line fluxes for [\oi]$\lambda$6300 and H$\gamma$ in the Prism spectrum. For each separate grating spectrum, we model the continuum with a simple polynomial and the emission features with Gaussian line profiles, tying the redshift, $z_{\rm spec}$, and the line full-width-half-maximum (FWHM) across transitions. This assumes that the emission lines all originate from and trace the same ionized gas in the star-forming region of the galaxy. We derive a spectroscopic redshift of $z_{\rm spec} = 5.943 \pm 0.001$ and a line ${\rm FWHM} = 218\pm 144$\,km\,s$^{-1}$. The best-fit line model is shown in Fig. \ref{fig:med_res_fit} and the derived line fluxes are summarized in Table \ref{table:linefluxes}. 

\begin{table}[h!]
\centering
\begin{adjustbox}{width=8.5cm}
\begin{tabular}{c c c c c} 
 \hline
   & PRISM & G140M & G235M & G395M \\ [0.5ex] 
 \hline\hline
Ly$\alpha$  & $92.8 \pm 1.0$ & $52.5 \pm 1.2$ & $-$ & $-$ \\
$\rm [CIV]\;\lambda$1548  & $-$  & $8.2 \pm 1.0 $ &  $-$ & $-$ \\
$\rm [CIV]\;\lambda$1550  & $33.6 \pm 2.1 $  & $4.3 \pm 1.0 $ &  $-$ & $-$ \\
$\rm [HeII]\;\lambda$1640  & $-$  & $1.6 \pm 1.0$ &  $-$ & $-$ \\ 
$\rm [OIII]\;\lambda$1666  & $-$  & $1.4 \pm 1.0$ &  $-$ & $-$ \\ 
$\rm [OII]\;\lambda\lambda$3728  & $-$  & $-$ &  $1.5 \pm 0.3$ & $-$ \\ 
$\rm H9$  & $-$  & $-$ &  $1.0 \pm 0.3$ & $-$ \\ 
$\rm [NeIII]\;\lambda$3869  & $18.2 \pm 0.8$  & $-$ &  $5.2 \pm 0.3$ & $-$ \\ 
$\rm [HeI]\;\lambda$3889  & $-$  & $-$ &  $2.0 \pm 0.3$ & $-$ \\
$\rm H\gamma$  & $-$  & $-$ &  $-$ & $6.6 \pm 0.2$ \\ 
$\rm [OIII]\;\lambda$4363  & $-$  & $-$  &  $-$  & $2.4 \pm 0.2$  \\ 
$\rm H\beta$  & $17.5 \pm 0.9$ & $-$ &  $-$  & $13.2 \pm 0.3$  \\ 
$\rm [OIII]\;\lambda$4959  & $32.8 \pm 0.05$ & $-$ & $-$ & $27.1 \pm 0.3$ \\ 
$\rm [OIII]\;\lambda$5007  & $88.7 \pm 0.39$  & $-$ &  $-$ & $88.7 \pm 0.3$ \\
$\rm [OI]\;\lambda$6300  & $0.5 \pm 0.2$ & $-$ &  $-$ & $-$  \\
$\rm H\alpha$  & $36.6 \pm 0.7$  & $-$ &  $-$ & $-$ \\
 [1ex] 
\hline
\end{tabular}
\end{adjustbox}
\caption[Line flux measurements reported in units of $\rm \times \; 10^{-19} \; erg \; s^{-1} \; cm^{-2}$]{Line flux measurements reported in units of $\rm \times \; 10^{-19} \; erg \; s^{-1} \; cm^{-2}$.}
\label{table:linefluxes}
\end{table}

\subsection{Star-formation rate and rest-frame UV characteristics}

The derived ratios of the three strongest Balmer lines, ${\rm H}\beta/{\rm H}\gamma = 2\pm 0.08$ and ${\rm H}\alpha/{\rm H}\beta = 2.77\pm 0.08$, are consistent with the theoretically predicted values assuming a Case B recombination scenario with $T_e = 10^{4}$\,K \citep{Osterbrock06}, indicating negligible dust attenuation. We caution, however, that H$\alpha$ is not covered by any of the grating spectra, so the ratio ${\rm H}\alpha/{\rm H}\beta$ is measured using the H$\alpha$ line flux derived from the Prism spectrum. We further integrate the spectral region covering H$\beta$ and the [\oiii]\,$\lambda \lambda 4960,5008$ doublet, yielding a joint equivalent width (EW) of $2266.4\pm 71.1\,\AA$. This is among the highest 16th percentile distribution of the full \jwst-PRIMAL sample, revealing an intensely star-forming galaxy. 

We derive the star-formation rate (SFR) following \citet{Kennicutt98} based on the H$\alpha$ line luminosity
\begin{equation}
    {\rm SFR}_{\rm H\alpha} \, (M_\odot \, {\rm yr}^{-1}) = 5.5 \times 10^{-42}L_{\rm H\alpha} \, ({\rm erg\,s}^{-1})
\end{equation}
assuming the initial mass function (IMF) from \citet{Kroupa01}. This is more top-heavy than the typical adopted Salpeter or Chabrier IMFs, and thus likely more appropriate for high-redshift galaxies \citep[e.g.,][]{Steinhardt23}. This yields an SFR of $8.20 \pm 0.15\,M_\odot\,{\rm yr}^{-1}$, supporting a significant star formation activity. Assuming a different IMF introduces an additional $\approx 30\%$ systematic uncertainty on the SFR, which we propagate to the SFR in the following analysis. We further note that the measured line width ($\approx 218$\,km\,s$^{-1}$) and the derived [\oiii]\,$\lambda 5007 / H\beta$ versus [\oi]\,$\lambda 6300/H\alpha$ line flux ratios do not indicate any robust signatures of an AGN based on the demarcation line from \citet{Kewley_2001}. This source may thus simply be a highly-ionized star-forming galaxy, as also supported by the high ionization parameter, $O_{32}$ = [\oiii]\,$5008$ / [\oii]\,$\lambda 3728$ = $59\pm 20$. We note, however, that this demarcation may not be applicable at high redshifts \citep{Ubler23,Calabro24}, and indeed propositions for an AGN origin based on the emission line flux and continuum has been made for this particular source \citep{Scholtz23,Tacchella24}.   

We infer the absolute UV magnitude of the source by integrating the flux density of the photometrically-calibrated spectra in the region around rest-frame $\approx 1500\,\AA$, yielding $M_{\rm UV} = -19.56\pm 0.05$\,mag or $L_{\rm UV} = (3.15\pm 0.16)\times 10^{28}$\,erg\,Hz\,s$^{-1}$. We derive the equivalent SFR from the UV luminosity again following \citet{Kennicutt98}
\begin{equation}
    {\rm SFR}_{\rm UV} \,(M_\odot \, {\rm yr}^{-1}) = 10^{-28} L_{\rm UV}\,({\rm \,erg\,Hz\,s^{-1}})
\end{equation}
here assuming 1/10 solar metallicity. This yields ${\rm SFR}_{\rm UV} = 3.2\pm 1.2\,M_\odot\,{\rm yr}^{-1}$, consistent within $2\sigma$ of the estimate from the H$\alpha$ recombination line. From the spectra, we can further derive the ionizing photon production efficiency of the source, $\xi_{\rm ion}$ \citep{Bouwens16}. Following \citet{Matthee23}, we derive
\begin{equation}
    \xi_{\rm ion}\, ({\rm Hz\, erg^{-1}}) = \frac{L_{\rm H\beta}\,({\rm erg\, s^{-1}})}{c_{\rm H\beta}({\rm erg}) ~ L_{\rm UV}({\rm erg\, s^{-1}\, Hz^{-1}})} 
\end{equation}
where $c_{\rm H\beta} = 4.86\times 10^{-13}$\,erg is the H$\beta$ line-emission coefficient, assuming a Case B recombination scenario with $T_e = 10^{4}$\,K and a zero escape fraction of ionizing Lyman Continuum (LyC) photons, $f^{\rm LyC}_{\rm esc} = 0$ \citep[e.g.,][]{Schaerer03}. This yields $\log \xi_{\rm ion} = 25.62\pm 0.02$, which is among the highest efficiency rates of galaxies at $z\gtrsim 6$ \citep[e.g.,][]{Matthee23,Atek23,Fujimoto23_uncover,Heintz24} and slightly above the canonical value on average required to ionize the IGM \citep{Robertson13}. 

Finally, we measure the spatial size of GS9422 in the F105W filter (rest-frame UV $\approx 1500\,\AA$) with a 2D Gaussian model, which provides a good match to the apparent simple morphology of the source (see Fig.~\ref{fig:introfig}). From the best-fit 2D Gaussian model we extract semi-major and -minor axes, $a'$ and $b'$, and derive an effective half-light UV radius $R_{\rm eff,UV} = \sqrt{a'b'} = 0\farcs 8$\, corresponding to a projected physical size of 0.59\,kpc at $z=5.943$. The intense SFR and small physical size, indicate that this source has an SFR surface density, $\log (\Sigma_{\rm SFR} / M_\odot\,{\rm yr}^{-1}\,{\rm kpc}^{-2}) = 1.14\pm 0.30$, among the highest known of the local galaxy population \citep[e.g.,][]{Kennicutt12}. 

\subsection{Gas-phase metallicity}

\begin{figure}
    \centering
    \includegraphics[width=9cm]{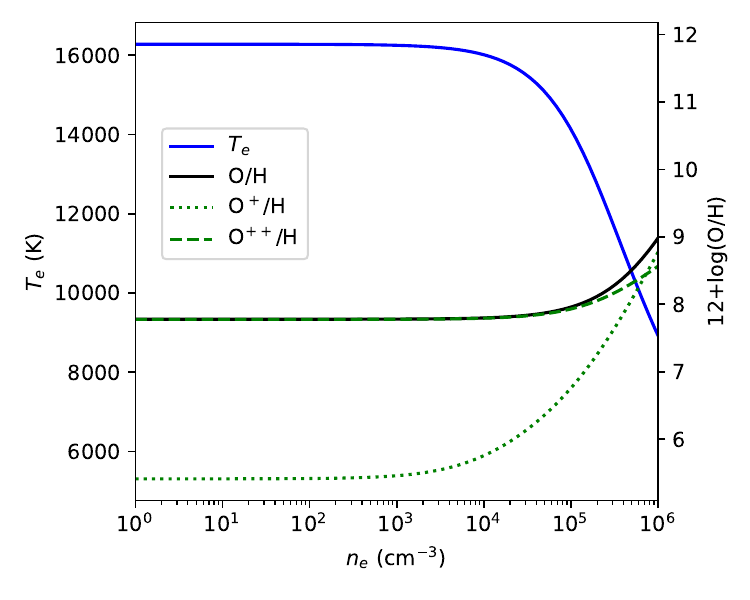}
    \caption{Electron temperature $T_e$ as a function of density, $n_e$ based on the observed nebular and auroral [\oiii], [\oii], and H$\beta$ line fluxes following the prescription by \citet{Izotov06}. The derived electron temperature is $T_e = (1.6\pm 0.15)\times 10^{4}$\,K indicating an oxygen abundance of $\rm 12+\log(O/H) = 7.78\pm 0.10$, for any electron densities $n_e < 10^{4} \; \rm cm^{-3}$. }
    \label{fig:telogoh}
\end{figure}

In the $\mathcal{R}\approx 1000$ grating spectrum we are able to detect and resolve the auroral line [\oiii]\,$\lambda 4363$ from H$\gamma$. This enables us to determine a temperature-sensitive estimate of the gas-phase metallicity of this galaxy (the so-called direct $T_e$-method) in combination with the nebular lines [\oiii]\,$\lambda\lambda 4960,5008$ and H$\beta$ emission lines. Following the iterations outlined in \citet{Izotov06} we infer an electron temperature $T_e = (1.6\pm 0.15)\times 10^{4}$\,K and oxygen abundance of $\rm 12+\log(O/H) = 7.78\pm 0.10$, for any electron densities $n_e < 10^{4} \; \rm cm^{-3}$ (common for most \hii\ regions), see Fig.~\ref{fig:telogoh}. This corresponds to an approx. $10\%$ solar metallicity \citep[for $\rm 12+\log(O/H)_\odot = 8.7$;][]{Asplund09}, common for galaxies at the same redshift \citep{Langeroodi23,Heintz23_FMR,Curti23} for the given SFR and stellar mass (see Sect.~\ref{ssec:sedfit} below). Our estimate is further consistent with that derived by \citet{Cameron_2023}.    

\begin{figure}
    \centering
    \includegraphics[width=8.5cm]{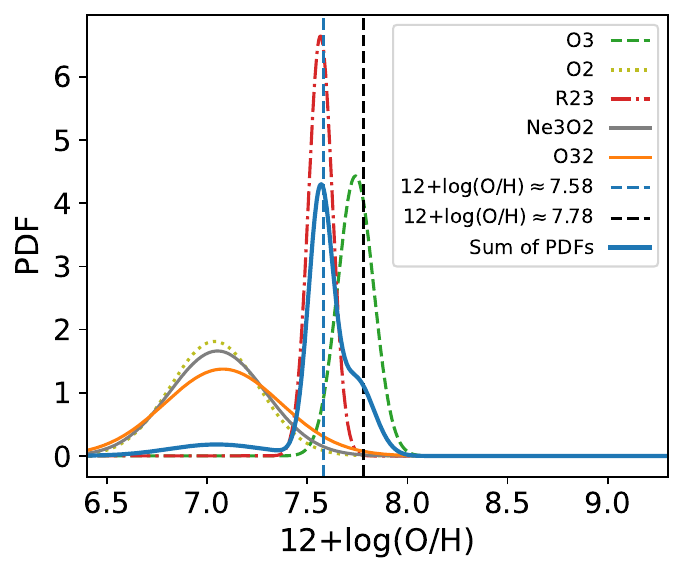}
    \caption{Normalized PDFs for each metallicity calculation. Blue and black vertical lines represent the $\rm 12+log(O/H)$ values calculated using the strong-line calibrations method by \citet{Sanders24} and $T_{e}$-method, respectively.}
    \label{fig:metallicity}
\end{figure}

We use the metallicity derived from the direct $T_e$-method to evaluate the various strong-line ratios inferred for galaxies at $z>6$ \citep[e.g.,][]{Nakajima23,Sanders24}. Specifically, we test the more conservative approach by \citet{Heintz24}, deriving the joint posterior distribution function (PDF) from each applicable strong-line diagnostic inversely weighted by the scatter for each relation and represent the final metallicity as the median and 16th to 84th percentiles of the joint PDF. We assume the single set of calibrations from \citet{Sanders24} and show the individual and joint PDFs in Fig.~\ref{fig:metallicity}. From this approach, we find a median $\rm 12+\log(O/H) = 7.58$ of the distribution, with 16th to 84th percentiles of 5.8 to 9.2. This is consistent within $1\sigma$ for the measurement from the direct $T_e$-method. We note that the $O_3$ calibration shows the best qualitative agreement with the direct $T_e$-method for this particular case, likely due to the high ionization parameter of the source. Assuming instead the $R23$ calibration, for instance, would underestimate the actual metallicity at more than $3\sigma$. This more conservative approach thus seems to optimally take into account the uncertainties related to each individual calibrator.

\subsection{Spectro-photometric SED modelling} \label{ssec:sedfit}

We perform joint spectro-photometric modelling of the \jwst/NIRSpec and NIRCam data to derive the spectral energy distribution (SED) of the source and extract additional physical parameters using \textsc{Bagpipes} \citep{Carnall_2018}. We modify the default \textsc{Bagpipes} framework to incorporate an additional \lya\ absorption component that allows us to additionally constrain the \hi\ column density in the SED model of the galaxy. This is particularly important for this galaxy as the DLA feature dominates the continuum flux at rest-frame UV wavelengths near the \lya\ emission line profile. The \lya\ optical depth from local ISM \hi\ gas is described through the Voigt-Hjerting absorption profile following the analytical approximation derived by \citet{TepperGarcia06} as shown in Eq. ~\ref{eq:abs_profile}. 
\begin{equation}
    \tau_{\rm ISM}(\lambda) = C\,a \, H(a,x) \, N_{\hi} 
    \label{eq:abs_profile}
\end{equation}
Here, $C$ is the photon absorption constant, a is the damping parameter, and $H(a,x)$ is the Voigt-Hjerting function. We further model the optical depth due to the Gunn-Peterson effect from an increasingly neutral IGM, given by \citet{Miralda_Escude_2000} with
\begin{equation}
    \tau_{\rm GP}(z) = 1.8\times 10^{5} h^{-1} (\Omega_{\rm DM,0})^{-1/2} x_{\rm HI} \left(\frac{\Omega_{\rm m,0} h^2}{0.02} \right) \left(\frac{1+z}{7} \right)^{3/2} 
\end{equation}
We modify this following the formalism of \citet{Totani_2006}
\begin{multline}
    \tau_{\rm IGM}(\lambda,z) = \frac{x_{\hi}R_{\alpha}\tau_{\rm GP}(z_{\rm spec})}{\pi} \left( \frac{1+z_{\rm abs}}{1+z_{\rm spec}}\right) ^{3/2}  
    \times 
    \\
    \left[ I\left( \frac{1+z_{\rm IGM,u}}{1+z_{\rm abs}} \right) - I\left( \frac{1+z_{\rm IGM,l}}{1+z_{\rm abs}} \right) \right]
    \label{eq:gunn-peterson_abs}
\end{multline}
where $z_{\rm abs}$ is the observed redshift of the neutral gas, $x_{\rm \hi}$ is the average neutral to total hydrogen fraction of the IGM (assumed to be 0.01), $R_{\alpha} = \Lambda_{\alpha}\lambda_{Ly\alpha}/(4\pi c) = 2.02 \times 10^{-8}$ is a dimensionless quantity that includes the damping constant of the \lya\ resonance ($\Lambda_{\alpha}$) and the \lya\ wavelength ($\lambda_{Ly\alpha}$). This correction assumes that the IGM is uniformly distributed within the redshift range $z_{\rm IGM,l}$ and $z_{\rm IGM,u}$. We set the upper bound at the galaxy redshift, $z_{\rm IGM,u} = z_{\rm spec}$, and the lower bound to $z_{\rm IGM,l} = 5$. We fix the \textsc{Bagpipes} model's spectroscopic redshift to $z_{\rm spec}$ (i.e. redshift of the galaxy) while keeping $z_{\rm abs}$ as a free parameter. We further assume a double-power-law star formation history and Calzetti dust attenuation.

First, we attempt to model the DLA assuming an absorption redshift, $z_{\rm abs}=z_{\rm spec}$. The result is presented in Appendix ~\ref{fig:SED1}. Fixing $z_{\rm abs}$ to the spectroscopic redshift does not fit the data satisfactorily. Consequently, we retain the absorption redshift as a free parameter.

\begin{figure*}
    \centering
    \includegraphics[width=14cm]{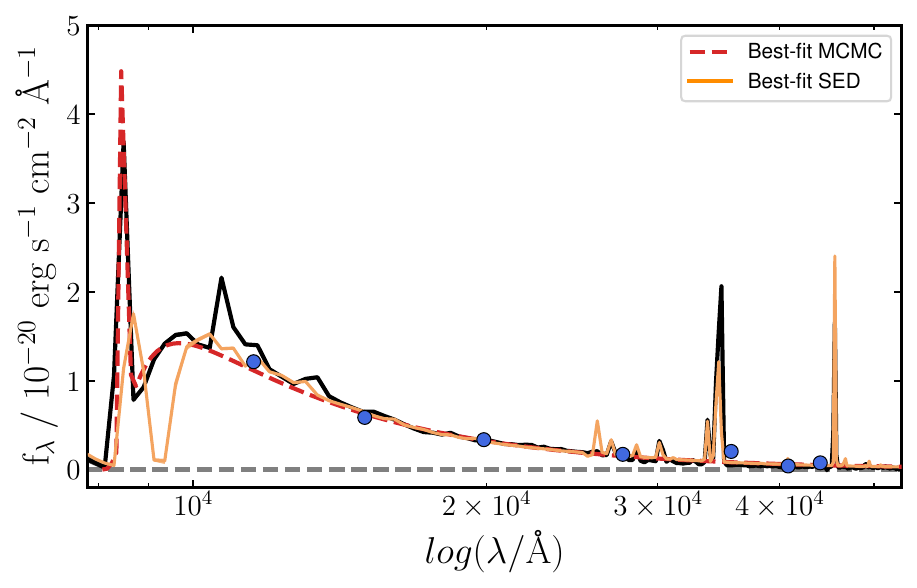}
    \caption{Best-fit spectro-photometric SED and \lya model of the \jwst data. The black solid curve shows the \jwst/NIRSpec Prism spectrum and the dark blue dots the corresponding \jwst/NIRCam photometry. The best-fit SED model from \textsc{Bagpipes} is shown as the orange curve, here modified to include the DLA, with $z_{\rm abs}$ as a free parameter. The best-fit MCMC stellar continuum and \lya\ model is shown as the red dashed line, with a best-fit \lya\ emission line redshift consistent with $z_{\rm spec}$ but $z_{\rm abs} = 5.396 \pm 0.098$. 
    }
    \label{fig:SED+MCMC}
\end{figure*}

The best-fit SED model matched to the available spectroscopic and photometric data, with $z_{\rm abs}$ as a free parameter, is shown in Fig.~\ref{fig:SED+MCMC}. From the fit, we derive a stellar mass ${\rm log}(M_\star/M_{\odot}) = 7.8 \pm 0.01$, ionization parameter ${\rm \log} U = -2.001^{+0.002}_{-0.001}$, and mass-weighted age of $\tau_{\rm mass} = (10.9^{+0.07}_{-0.12})$\,Myr. The galaxy falls slightly above the star-forming main-sequence (SFMS) of galaxies at $z\approx 6$ \citep{Thorne21,Heintz23_FMR}. However, the stellar mass may be underestimated by 0.5\,dex in our SED modelling due to "outshining" effects \citep{Whitler23,GimenezArteaga23,Narayanan24} concealing the more massive, older stellar population, which would place it consistently on the SFMS. As is clear from Fig.~\ref{fig:SED+MCMC} we are not able to accurately constrain the damping as the large \hi\ column density implied by the width of the damping wings is inconsistent with the rollover and absorption trough redshift. This was also noted by \citet{Cameron_2023}, motivating their modelling of the \lya\ rollover as produced by two-photon emission. Here, we instead examine the parameter space first assuming the most simple DLA scenario as outlined in Sect.~\ref{ssec:dla} below.

\begin{table}[h!]
\centering
\begin{adjustbox}{width=7.5cm}
\begin{tabular}{@{}lcr@{}}
 \hline\hline
 R.A. (J2000) && $03^{\rm h}32^{\rm m}29.2^{\rm s}$ \\
 Decl. (J2000) && $-27^\circ 47' 51.48''$ \\
$z_{\rm spec}$ & & $5.943 \pm 0.001$ \\
$R_{\rm e} \; [kpc]$ & & $0.59$ \\
$A_{\rm e} \; [kpc^{2}]$ & & $1.85$ \\
$M_{\rm UV} \; [{\rm mag}]$ & & $-19.56\pm 0.05$ \\
$L_{\rm UV} \; [{\rm erg\, s^{-1}\, Hz^{-1}}]$ & & $(3.15\pm 0.16)\times 10^{28}$ \\
\\
$\beta_{\rm UV}$ & & $-2.36\pm 0.10$ \\
$\rm [\oiii]+H\beta ~ EW \; [\AA]$ & & $2266.4\pm 71.1$ \\
$\rm{SFR}_{\rm H\alpha} \; [M_{\odot} \; yr^{-1}]$ & & $8.2 \pm 2.8$ \\
$ \log \xi_{\rm ion}\; [{\rm Hz\, erg^{-1}}]$ & & $25.62\pm 0.02$ \\
$^{a}\rm 12 + log(O/H)$ & & $7.58^{+1.62}_{-1.78}$ \\
$\log(\rm Z/Z_{\odot})$ & & $-1.13 \pm 0.05$ \\
$T_e \, \rm [K]$ & & $(1.6\pm 0.15)\times 10^{4}$ \\
$f_{\rm esc,Ly\alpha}$ & & $0.3 \pm 0.2$ \\
\\
$^{c}\log (M_\star / M_\odot) $ & & $7.80 \pm 0.01$ \\
$^{c}\tau_{\rm mass}$ [Myr] & & $10.9^{+0.07}_{-0.12}$ \\
$^{c}\log U$ & & $-2.001^{+0.002}_{-0.001}$ \\
$^{c} A_{V}$ [mag] & & $0.041 \pm 0.013$ \\
\\
$^{b}z_{\rm abs}$ & & $5.396 \pm 0.098$ \\
$^{b}\log(N_{\rm \hi}) \; [cm^{-2}]$ & & $23.68 \pm 0.10$ \\ 
$\log\Sigma_{\rm SFR} \; [M_{\odot} \; yr^{-1} \; kpc^{-2}]$ & & $1.14 \pm 0.30$ \\ 
$\log\Sigma_{\rm gas} \; [M_{\odot} \; pc^{-2}]$ & & $3.01 \pm 0.01$ \\
$^{d}\log(N_{\rm \hi}) \; [cm^{-2}]$ & & $22.94 \pm 0.01$ \\ 
$^{d}M_{\rm \hi} \; [M_{\odot}]$ & & $(1.27 \pm 0.03) \times 10^{9}$\\
$^{d}t_{\rm \hi, depl.} \; [Gyr]$ & & $1.55 \pm 0.53$\\
\hline
\end{tabular}
\end{adjustbox}
\caption[Determined and measured physical properties of GS9422]{Determined and measured physical properties of GS9422 (without values measured with \textsc{Bagpipes}). $^{a}$Derived following the procedure in \cite{Sanders24}. $^{b}$From MCMC samling. $^{c}$Derived from the SED modelling with \textsc{Bagpipes}. $^{d}$Based on the Kennicutt-Schmidt law.}
\label{table:measurements}
\end{table}

\subsection{\lya emission and DLA modelling} \label{ssec:dla}

To model the continuum near the \lya\ edge for an accurate measure of the DLA and \lya\ emission feature, we describe the rest-frame UV spectral shape with the model
\begin{multline}
       F (\lambda) = F_{\rm 0} \lambda^{\beta_{\rm UV}} ~ \times ~ \tau_{\rm ISM}(\lambda) ~ +~ \tau_{\rm IGM}(\lambda,z) ~\times ~ \\ \frac{f_{\rm Ly\alpha}}{\sqrt{2\pi}\sigma_{\rm Ly\alpha}} \exp{\left( \frac{-(\lambda - \lambda_{\rm Ly\alpha,peak})^2}{2\sigma^2}\right)} 
\end{multline}
We find $\beta_{\rm UV} = -2.36\pm 0.10$, and assume this as the intrinsic continuum spectrum of the source. We superimpose the DLA feature and \lya emission line described by a Gaussian on this model, assuming a prior on $z_{\rm Ly\alpha,em}=z_{\rm spec}$ and leave the DLA redshift, $z_{\rm abs}$, as a free parameter. In the fit, we mask out the two strongest identified UV emission lines from [\civ] and [\ciii]. We use MCMC through the {\tt emcee} package to sample the posterior distribution of the modeled parameters. Cornerplots of the full set of posterior distributions are shown in Appendix, Fig.~\ref{fig:mcmc} and ~\ref{fig:mcmc2}.
We find that the redshift inferred for the \lya\ emission is consistent with that derived from the longer-wavelength Balmer recombination lines, and fix $z_{\rm Ly\alpha,em}=z_{\rm spec}$ in the next iteration. The model parameter for the DLA redshift is well converged on $z_{\rm abs} = 5.396 \pm 0.098$, with the full posterior distribution shown in Fig.~\ref{fig:cluster}. If this lower redshift solution for the DLA is robust, this would naturally explain the simultaneous presence of a DLA feature and strong \lya\ emission, which would be significantly outside the resonance frequency to not be entirely absorbed. We discuss the potential implications and provide a physical scenario in support of the lower-$z$ solution in Sect.~\ref{sec:disc}. 

We measure the \lya\ escape fraction as
\begin{equation}
    f_{\rm esc,Ly\alpha} = \frac{F_{\rm obs,Ly\alpha}}{8.7\times F_{\rm obs,H\alpha}}
\end{equation}
assuming an intrinsic ${\rm Ly}\alpha/{\rm H}\alpha$ ratio of 8.7 from the Case B recombination model and the inferred negligible dust component. This yields $f_{\rm esc,Ly\alpha} = 0.3\pm 0.2$, i.e. $\approx 30\%$, consistent with the general trend for faint ($M_{\rm UV} < -19.5$\,mag) galaxies at similar redshifts \citet{Saxena23}. This particular source thus likely represents the galaxy population that drives the large-scale reionization.

\section{Interpreting the strong Lyman-$\alpha$ absorption and emission} \label{sec:disc}

To disentangle the potential contributions to the strong observed \lya\ damping wings we first consider multiple scenarios. First, the two-photon emission model presented by \citet{Cameron_2023} appears to well match the rest-frame UV data and naturally explains the strong \lya\ emission line as well. However, given that the derived electron temperatures and gas-phase metallicities are an order of magnitude lower or higher, respectively, than required by this model we argue that this scenario is unlikely \citep[as also argued by][]{Chen23}. Second, an AGN may produce large UV-ionized proximity zones and thereby allow a substantial fraction of \lya\ photons to escape, but since all our measurements are consistent with a typical star-forming galaxy at $z\sim 6$ we instead pursue the foreground DLA solution. 

To disentangle the potential contributions to the observed \lya\ damping wing, we first consider the SFR surface density which we measured to be $\log (\Sigma_{\rm SFR} / M_\odot\,{\rm yr}^{-1}\,{\rm pc}^{-2}) = 1.14\pm 0.30$. This corresponds to a gas surface density, $\log (\Sigma_{\rm gas}/M_\odot {\rm pc}^{-2}) = 3.01\pm 0.30$, assuming the locally-derived Kennicutt-Schmidt (KS) relation \citep{Kennicutt98,Heiderman10,Kennicutt12}. Converting this into a gas mass and assuming that $M_{\rm \hi}=2/3 \, M_{\rm gas}$ (where the molecular gas mass constitutes $M_{\rm H_2} = 1/3 \, M_{\rm gas}$) at $z\approx 6$ \citep{Heintz22}, this yields $M_{\rm \hi} = 2/3 \, \Sigma_{\rm gas} A_e = (1.27\pm 0.03)\times 10^{9}\,M_\odot$. This corresponds to an integrated \hi\ column density of $N_{\rm \hi} = M_{\rm \hi} / (m_{\rm \hi} A_e) = (8.58\pm 0.02)\times 10^{22}\,$cm$^{-2}$, where $m_{\rm HI}$ is the mass the hydrogen atom. This is an order of magnitude lower than the \hi\ column density inferred from the strength of the \lya\ damping wing and implies an \hi\ gas mass fraction of $M_{\rm \hi} / M_\star = 0.20 \pm 0.01$ and a gas depletion time, $t_{\rm depl,\hi} = M_{\rm \hi} / {\rm SFR_{H\alpha}} = 1.55 \pm 0.53\,$Gyr. This supports an additional contribution to the integrated \hi\ column density or potentially that the typical gas masses are underestimated by $\approx 10\times$ in galaxies at $z\approx 6$ for a given SFR density compared to the locally-calibrated KS relation. We argue that this is unlikely since this relation has been robustly validated up to $z\gtrsim 2$ \citep[e.g.,][]{Danielson11} and would imply an \hi\ gas mass fraction, $M_{\rm \hi} / M_\star \approx 20$ for this particular source. 

\begin{figure*}
    \centering
    \includegraphics[width=15cm]{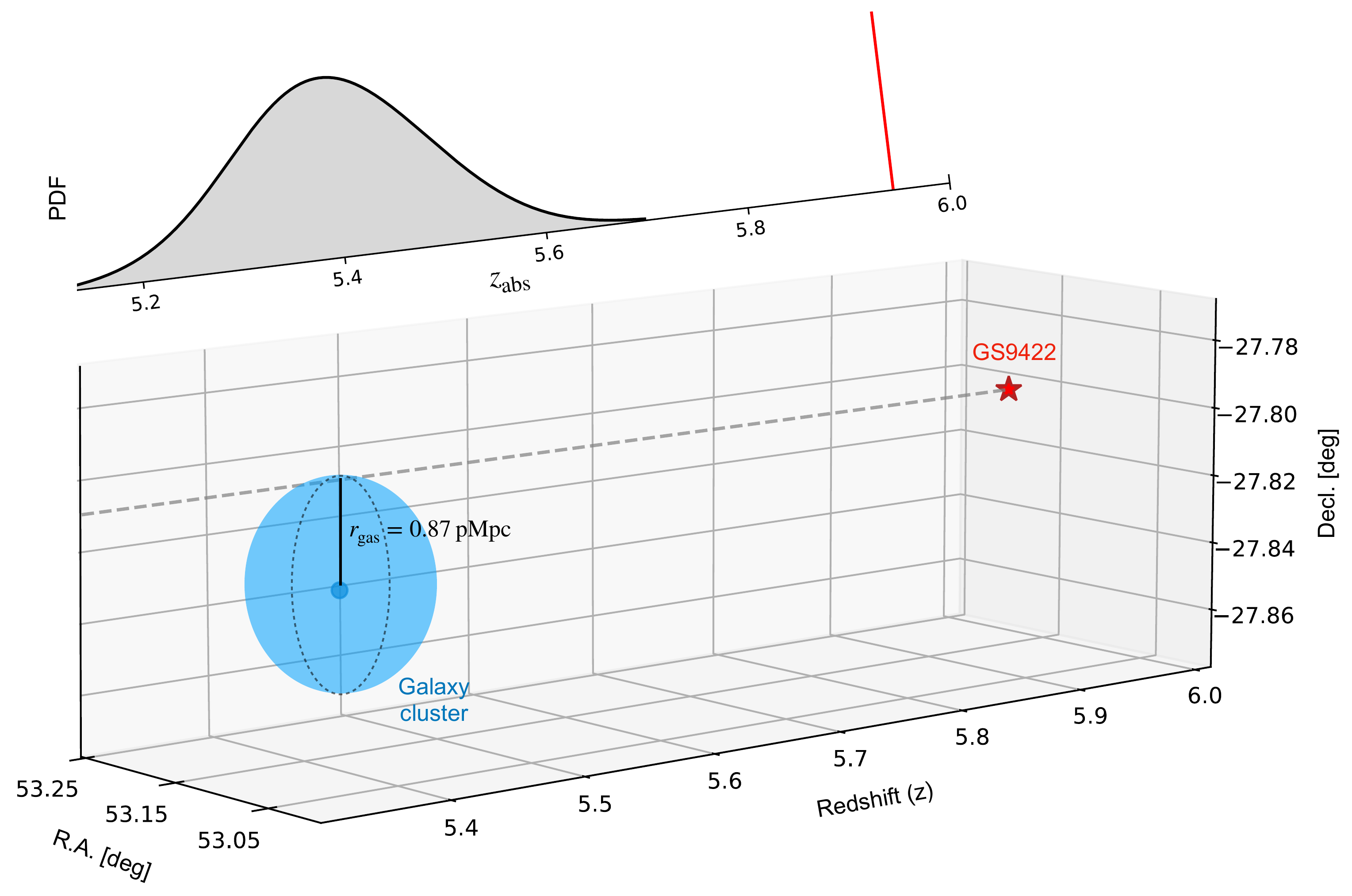}
    \caption{3D illustration of the proposed scenario with GS9244 at $z=5.943$ being background to a galaxy-cluster overdensity at $z=5.4$. The estimated center and mean redshift of the cluster is marked by the small blue dot \citep{Helton23}, and the blue sphere with radius $r=0.87$\,pMpc shows the impact parameter to the galaxy sightline. The smoothed probability density function (PDF) from the MCMC analysis (Fig.~\ref{fig:mcmc2}) is shown at the top, distributed around the cluster-redshift with median $z=5.4$.   
    }
    \label{fig:cluster}
\end{figure*}

Instead, the most viable solution is that the DLA traces a dense \hi\ gas reservoir in the foreground of the galaxy at $z = 5.40\pm 0.10$. This would also naturally allow for substantial escaping \lya\ emission as the resonance frequency of the transition is equally shifted to higher frequencies. Serendipitously, we note that there has been reported a massive galaxy overdensity at exactly $z = 5.4$ in nearby projection to the source presented here \citep{Helton23}. The impact parameter to the center of the proposed galaxy proto-cluster corresponds to $\approx 0.9$\,pMpc at the cluster redshift, as illustrated in Fig.~\ref{fig:cluster}. This is well within typical galaxy clusters such as Coma and Virgo observed in the local Universe and indeed includes most of the cluster members at $z=5.4$ \citep{Helton23}. Intriguingly, this may imply that we are probing the dense, cold circumcluster medium (CCM) feeding the formation of the galaxy proto-cluster at $z=5.4$. \citet{Chen23} also presents circumstantial evidence for a similar scenario based on the presence of strong DLAs in a small set of galaxies near a galaxy overdensity at $z\approx 7.8$. However, the implied gas column and total \hi\ gas mass for such a structure is immense since the CCM is typically observed to be comprised of more diffuse gas. An alternative scenario could be that the measured \hi\ column density is the cumulative representation of multiple DLAs along the line of sight from the galaxy redshift to the foreground absorber \citep[see e.g.][]{Witstok24}. To fully corroborate this scenario and potentially measure the covering fraction and total CCM \hi\ mass we need larger statistical samples and sightlines at different impact parameters through these clusters.    

\section{Conclusions}

We have presented a comprehensive characterization of the stellar and gas-phase content of a galaxy near the end of the reionization era showing an anomalous superposition of both a strong \lya\ emission line and a broadened damped \lya\ absorption wing. This particular source had been observed as part of the JADES GTO survey, for which medium-resolution ($\mathcal{R}\approx 1000$) \jwst/NIRSpec grating and Prism-mode spectra and multi-band NIRCam imaging have been obtained. From the multitude of detected nebular emission lines, we measured the redshift to be $z_{\rm spec} = 5.943\pm 0.001$ and found the source to be actively forming stars at ${\rm SFR} = (8.20\pm 2.8)\,M_\odot\, {\rm yr^{-1}}$ with an \lya\ escape fraction of $f_{\rm esc,Ly\alpha} = 30\%$. 
Based on the detection of the auroral [\oiii]\,$\lambda 4363$ emission line, we further determined the electron temperature of the \hii\ region to be $T_e = (1.6\pm 0.15)\times 10^{4}\,$K which allowed us to accurately measure the oxygen gas-phase abundance to be $\rm 12+\log(O/H) = 7.78\pm 0.10$ via the direct $T_e$-method. This source thus appears as a typical, albeit highly intense, and star-forming galaxy at $z\approx 6$. 

To model the observed SED and \lya\ emission and broad DLA trough, we modified the default capabilities of the SED fitting tool {\sc Bagpipes} \citep{Carnall_2018} to additionally include a prescription for the \lya\ transmission for a given \hi\ column density, $N_{\rm HI}$. This yielded a dust-poor ($A_V = (0.041 \pm 0.013)$\,mag), young (age $\approx 11 \, {\rm Myr}$) star-forming galaxy with a stellar mass of ${\rm log}(M_\star/M_{\odot}) = 7.8 \pm 0.01$. The posterior distributions suggested an abundant \hi\ column density with $N_{\rm HI} > 10^{23}\,$cm$^{-2}$ though were unable to provide a good match to the data. Instead, we modelled the \lya\ region independently using an MCMC to sample the posterior distribution of the absorber. We found a lower-redshift solution $z_{\rm abs} = 5.40 \pm 0.10$, inconsistent with the galaxy redshift at $>3\sigma$ confidence, that qualitatively provided a much better fit the to data.

We demonstrated via the Kennicutt-Schmidt relation and based on the measured SFR surface density, $\log (\Sigma_{\rm SFR} / M_\odot\,{\rm yr}^{-1}\,{\rm kpc}^{-2}) = 1.14\pm 0.30$, that the galaxy itself is likely only contributing to $\lesssim 10\%$ of the integrated \hi\ gas column along the line of sight. This was further evidence that the majority of the \hi\ gas is located in the foreground and not associated physically with the galaxy. Intriguingly, a massive galaxy overdensity has recently been reported by \citet{Helton23}, located at $z = 5.387$ and in nearby ($\approx 1\,$Mpc) projected distance to the galaxy studied here. This scenario would also naturally resolve the simultaneous detection of strong \lya\ emission and abundant, self-shielding \hi\ columns probed via the DLA feature. While larger statistical samples and sightlines are needed to validate this scenario, this interpretation is intriguing and provides some ambiguity to the proposed two-photon nebular emission \citep{Cameron_2023} or the AGN+young stellar disk interpretations as the dominant contribution to the rest-frame UV spectral continuum.

\begin{acknowledgements}
We would like to thank Harley Katz for enlightening discussions on the physical interpretation of the two-photon emission model. 
This work has received funding from the Swiss State Secretariat for Education, Research and Innovation (SERI) under contract number MB22.00072.
The Cosmic Dawn Center (DAWN) is funded by the Danish National Research Foundation under grant DNRF140.
The data products presented herein were retrieved from the DAWN \jwst Archive (DJA). DJA is an initiative of the Cosmic Dawn Center, which is funded by the Danish National Research Foundation under grant DNRF140.

This work is based in part on observations made with the NASA/ESA/CSA James Webb Space Telescope. The data were obtained from the Mikulski Archive for Space Telescopes (MAST) at the Space Telescope Science Institute, which is operated by the Association of Universities for Research in Astronomy, Inc., under NASA contract NAS 5-03127 for \jwst. 
\end{acknowledgements}

%
%

\bibliographystyle{aa}
\bibliography{ref}

\begin{appendix}
\onecolumn

\section{Posterior distributions}

\begin{figure*}[!ht]
\centering
\includegraphics[width=18cm]{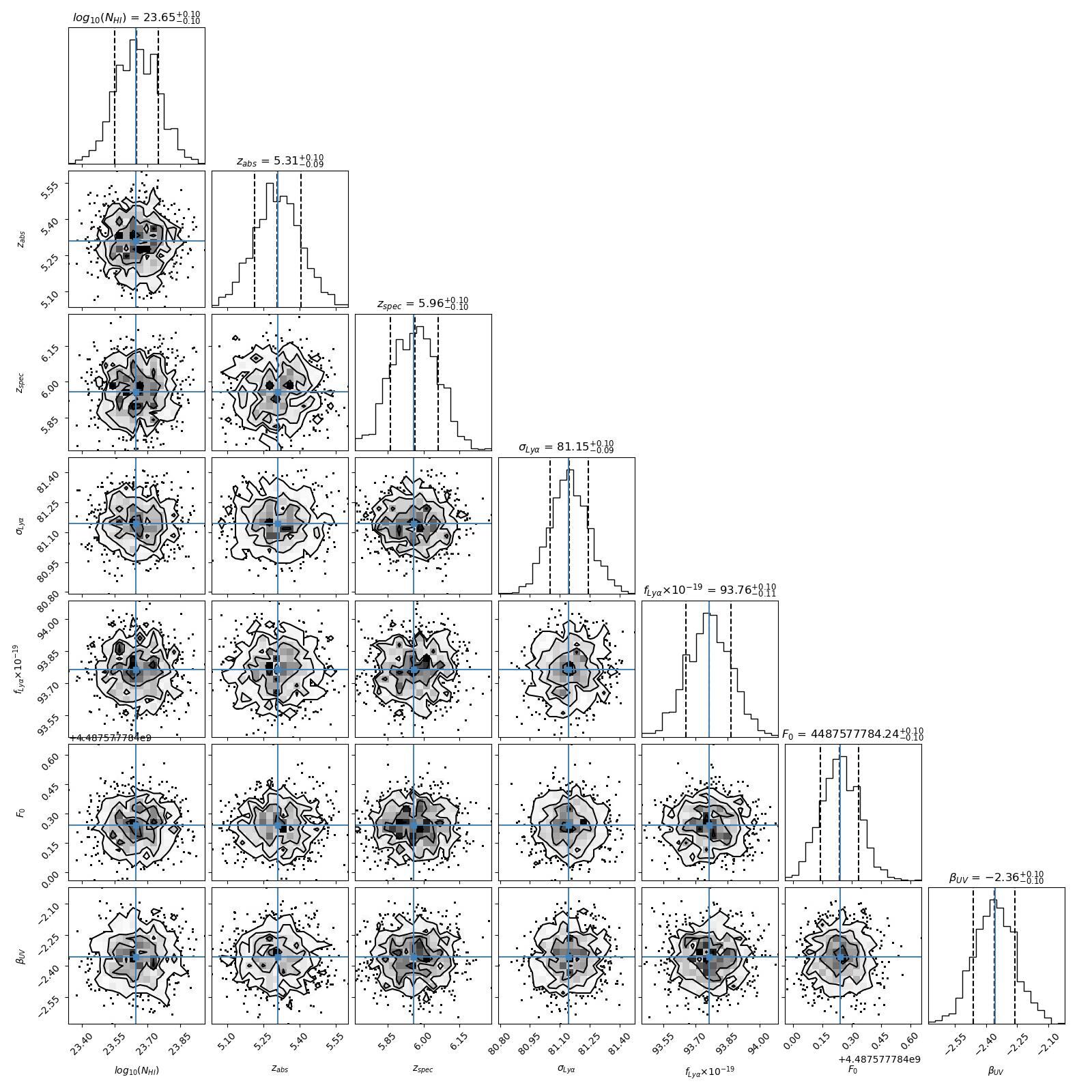}
\caption{Cornerplot for MCMC fit with \textit{no} fixed parameters.}
\label{fig:mcmc}
\end{figure*}

\newpage

\begin{figure*}[!ht]
\centering
\includegraphics[width=18cm]{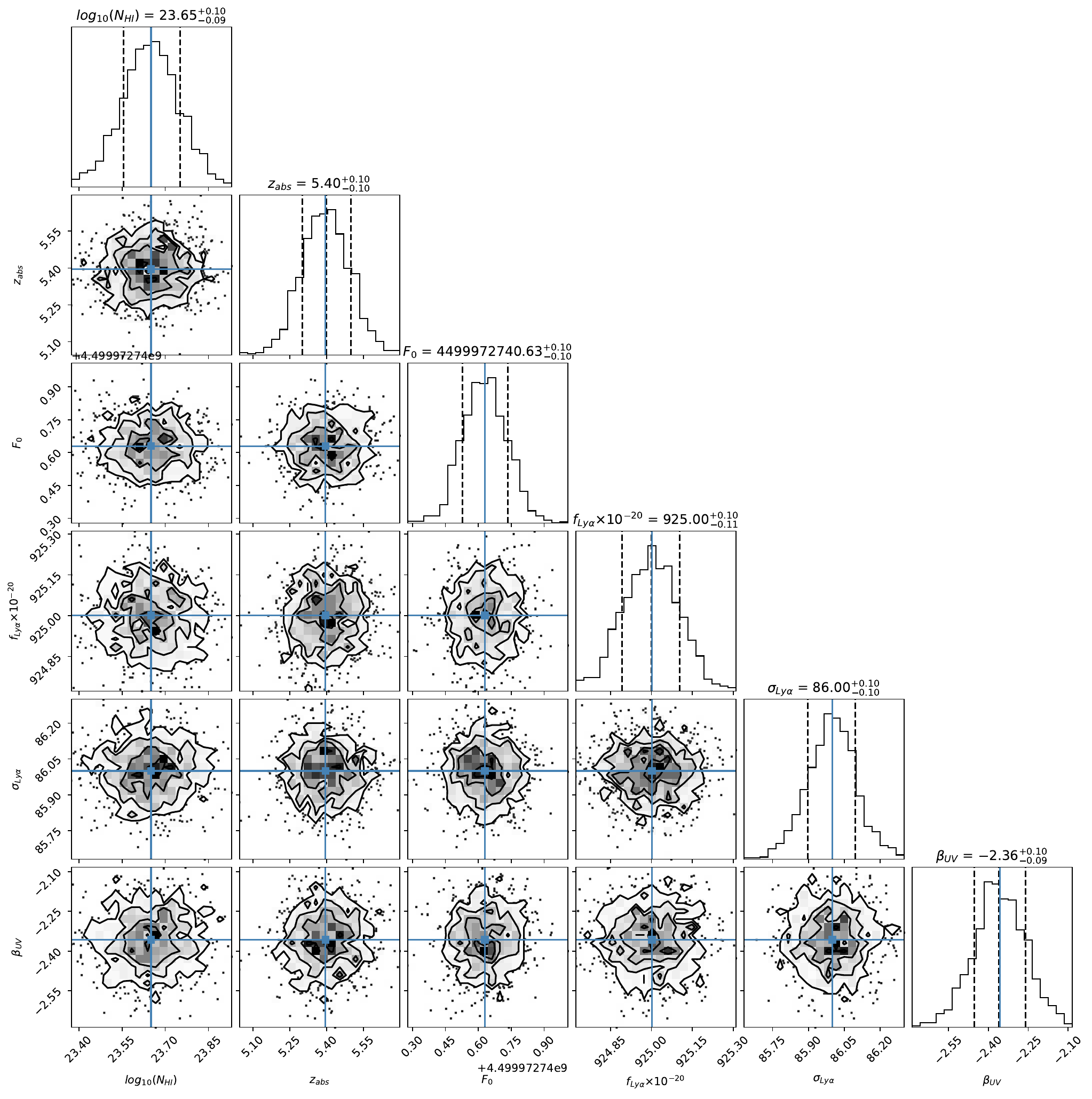}
\caption{Cornerplot for MCMC fit with $z_{\rm Ly\alpha,em} = z_{\rm spec}$ (fixed) and $z_{\rm abs}$ as a free parameter.}
\label{fig:mcmc2}
\end{figure*}

\newpage

\section{SED fitting with $z_{\rm abs} = z_{\rm spec}$}

\begin{figure*}[!ht]
    \centering
    \includegraphics[width=18cm]{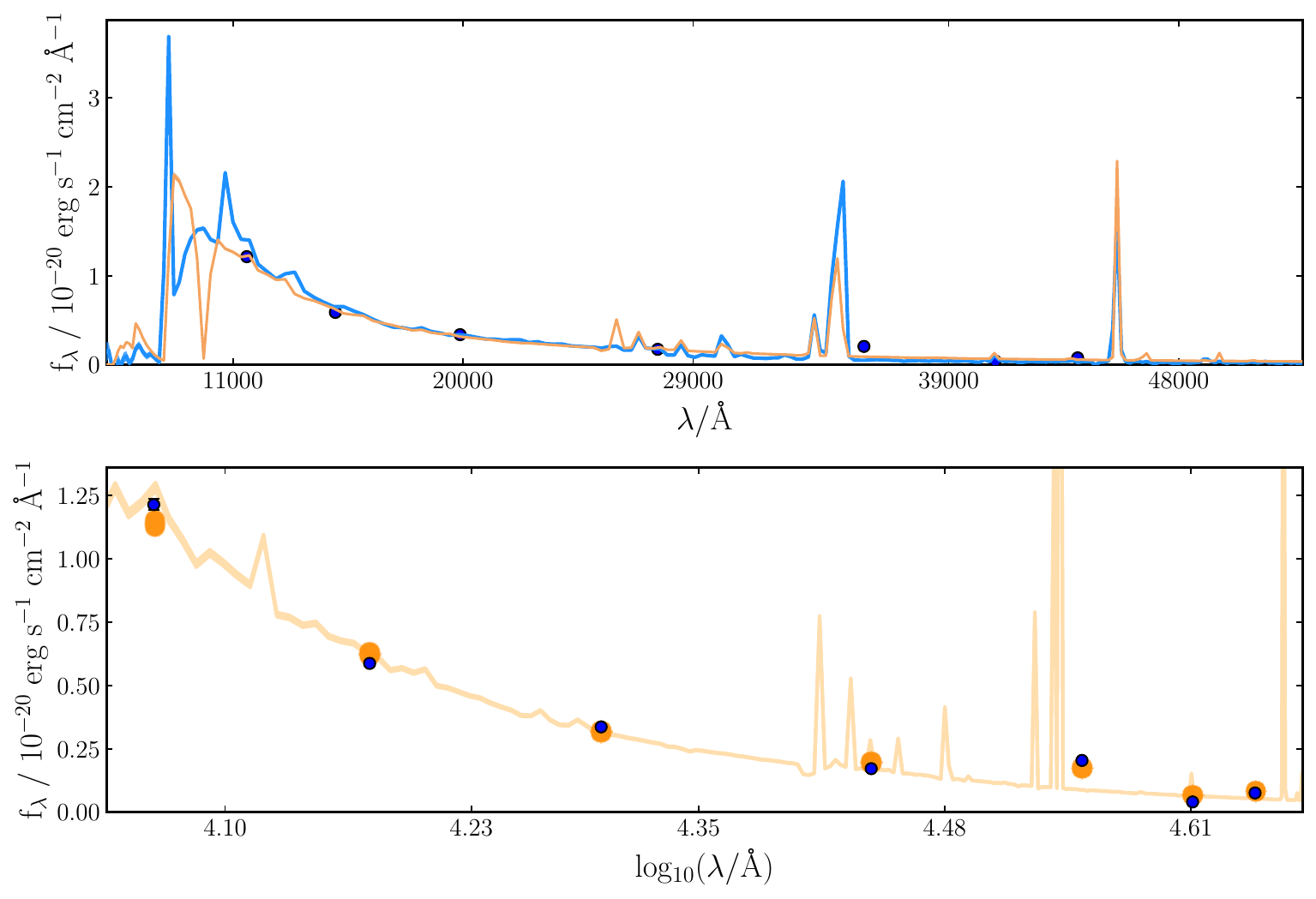}
    \caption{Best-fit SED model matched to both spectroscopic and photometric data with $z_{\rm abs} = z_{\rm spec}$.}
    \label{fig:SED1}
\end{figure*}

\end{appendix}

\end{document}